\documentclass[a4paper,11pt]{article}
\usepackage{jcappub}
\usepackage{graphicx,epsfig,color,times,wasysym}
\usepackage{amsmath}
\usepackage{epsfig}
\usepackage{anyfontsize}

\usepackage[table]{xcolor}
\usepackage{floatrow}
\newfloatcommand{capbtabbox}{table}[][\FBwidth]
\usepackage{capt-of}

\def\apj{Astrophys. J.}

\def\apjs{Astrophys. J. Suppl.}

\def\ijmpd{Int. J. Mod. Phys. D}

\begin{document}
\title{Consistency tests for Planck and WMAP in the low multipole domain}
\author{A. Frejsel, M. Hansen and H. Liu}  
\affiliation{Niels Bohr Institute and Discovery Center, Blegdamsvej 17, 2100 Copenhagen {\O}, Denmark }
\emailAdd{frejsel@nbi.dk}
\emailAdd{kirstejn@nbi.dk}
\emailAdd{liuhao@nbi.dk}

\abstract{
Recently, full sky maps from Planck have been made publicly available. In this paper, we do consistency tests for the three Planck CMB sky maps. We assume that the difference between two maps represents the contributions from systematics, noise, foregrounds and other sources, and that a precise representation of the Cosmic Microwave Background should be uncorrelated with it. We investigate the cross correlation in pixel space between the difference maps and the various Planck maps and find no significant correlations, in comparison to 10000 random Gaussian simulated maps. Additionally we investigate the difference map between the WMAP ILC 9 year map and the ILC 7 year map. We perform cross correlations between this difference map, and the ILC9 and ILC7, and find significant correlations only for the ILC9, at more than the $99.99\%$ level. Likewise, a comparison between the Planck NILC map and the WMAP ILC9 map, shows a strong correlation for the ILC9 map with the difference map, also at more than the $99.99\%$ level. Thus the ILC9 appears to be more contaminated than the ILC7, which should be taken into consideration when using WMAP maps for cosmological analyses.}

\maketitle

\section{Introduction}
The Cosmic Microwave Background (CMB) is a key source of information on the early Universe. The newly released Planck data \cite{Planck1} as well as data from WMAP \cite{WMAP9} play important roles in this study. Whereas Planck greatly improves the current CMB measurement precision, it will still be very interesting and relevant to compare CMB products obtained with Planck and WMAP respectively, like the WMAP ILC maps and the Planck full sky maps. 

The recent Planck CMB data release included the derived CMB products NILC, SMICA and SEVEM maps. The three maps are made via different techniques: the NILC map is a needlet variant of the Internal Linear Combination technique, SMICA is made via spectral parameter fitting in the harmonic domain, and SEVEM is constructed through template fitting using the lowest and highest frequency bands (see \cite{Planck12} for details). It is highly important to understand the differences and similarities of these three maps, when making cosmological conclusions.

Recently, the 9 year CMB data from WMAP has also been released, including the 9yr Internal Linear Combination map (ILC9) \cite{WMAP9}. The ILC9 is constructed through the combination of data at different frequencies with different weighting coefficients, in order to obtain a full sky map of the CMB with as little contamination (foregrounds, systematics and noise) as possible. 

The aim of this paper is twofold. We wish to test the internal consistency and differences between the three released Planck maps (SMICA, SEVEM and NILC) and the differences between the WMAP ILC9 and 7 maps (7 year WMAP ILC data \cite{WMAP7}). Furthermore, we wish to test the external consistency between the WMAP ILC9 map and the Planck NILC map.

The basic assumption of this paper is that the difference between two maps (the 'difference map') is composed of non-CMB contributions (such as noise, systematic errors, foregrounds etc.) that ought to be uncorrelated with the true, primordial CMB signal. If the maps do contain remnants of these contaminants they will be correlated with the difference map, the significance of which can be tested by comparing to simulations.

\section{Testing the map of errors}
Since we are faced with several different sky maps, all in principle depicting the CMB signal, we must take under consideration that they contain some element of contamination. Thus, the three Planck maps as well as the WMAP ILC maps are likely not perfect representations of the true primordial CMB, but contain an intrinsic CMB component, $c$, and a small non-cosmological component, $n$, which is due to noise, foregrounds, systematics etc. If we subtract two maps we will be left with a difference map, $d$, which is only composed of a difference in contaminants, $\Delta n$. In a spherical harmonics decomposition we can write this as
\begin{eqnarray}
	&&a^{map}_{lm}=c_{lm}+n_{lm} \nonumber \\
	&&d_{lm}=a^{map1}_{lm}-a^{map2}_{lm}=\Delta n_{lm},
	\label{eq:alm+noise}
\end{eqnarray}
where $a^{map}_{lm}$ refers to a specific map (WMAP or Planck). If the map is clean of contaminants (i.e. $n_{lm}^{map1}=0$) we would not expect a significant correlation between $a^{map1}_{lm}$ and $d_{lm}$, because a pure CMB signal of course should not be correlated significantly with noise, systematics or foreground residuals. One can consider a cross correlation coefficient in multipole space $K$, where we compare a difference map, $\Delta n_{lm}$, and one of the maps it was created from ($a_{lm}$). When we substitute via Eq. \ref{eq:alm+noise}, we get the following
\begin{eqnarray}
K(l)&=&\frac{\sum\limits_{m}(c_{lm}+n_{lm}^{a})\Delta n_{lm}^{*} + (c_{lm}+n_{lm}^{a})^{*}\Delta n_{lm}} {2(\sum\limits_{m}|c_{lm}+n_{lm}^{a}|^{2} \sum\limits_{m}|\Delta n_{lm}|^{2})^{1/2}} \nonumber \\
&=& \xi_{1} \left( \frac{\sum\limits_{m}|c_{lm}|^{2}}{\sum\limits_{m}|c_{lm}+n_{lm}^{a}|^{2}} \right)^{1/2} + \xi_{2} \left( \frac{\sum\limits_{m}|n_{lm}^{a}|^{2}}{\sum\limits_{m}|c_{lm}+n_{lm}^{a}|^{2}} \right)^{1/2}
\label{eq:cross_large}
\end{eqnarray}
where a * indicates a complex conjugate and $n_{lm}^{a}$ is the noise inherent in map $a$. $\xi_{1}$ is the cross correlation between the primordial CMB ($c_{lm}$) and the difference map ($\Delta n_{lm}$), and is expected to be at the order of $0$. $\xi_{2}$ is the cross correlation between the noise of map $a$ ($n_{lm}^{a}$) and the difference map, which is at the order of $1$ (we expect the noise to correlate strongly with the difference map). Term 1 in Eq. \ref{eq:cross_large} is thus on the order of $0$, while the first part of the second term ($\xi_{2}$) is of order $1$. Thus the interesting term in the equation, the last part of the second term, is essentially a noise to signal ratio and is the term one would effectively test via cross correlations. The case is the same if one performs the test in pixel space.

We estimate the significance of the correlations by comparing the correlation between the sky maps and their respective difference map with the correlation between randomly simulated CMB maps and the difference map. We therefore compare the cross correlations with 10000 random Gaussian simulations based on the $\Lambda$CDM theoretical power spectrum.

We calculate the cross correlation using the following definition of the cross correlation coefficient in pixel space:
\begin{equation}
	K_p = \frac{\sum\limits_{i} (x_i - \bar{X})(y_i - \bar{Y})}{\sqrt{(\sum\limits_{i} (x_i-\bar{X})^2)(\sum\limits_{i} (y_i-\bar{Y})^2)}},
\label{eq:crossp}
\end{equation}
where $K_p$ is the total cross correlation coefficient, $x_i$ and $y_i$ are the values of pixel $i$ for the two maps respectively and $\bar{X}$ and $\bar{Y}$ are the mean pixel values for the two maps. We know that some residuals of the galactic plane are definitely still present in the maps. We therefore mask out the galaxy using the WMAP KQ85 9yr mask \cite{WMAP9}, in order to see the effect of contaminants in the rest of the map. Thus, the sum in $i$ is only over unmasked pixels. Note that the sign of the correlation coefficient is not significant, since the choice of which map to subtract from the other is interchangeable. As such, a significant anti-correlation would change to correlation if the two maps were subtracted from each other in the opposite order.

Also, bear in mind that the power of the difference map is not very important for this investigation. What we are truly testing here is the morphology of the maps. We are comparing the same spot in the two maps (pixel $i$), and the estimator $K_p$ is normalized. Thus we are effectively looking at normalized pixel values between $-1$ and $1$, and the estimator $K_p$ shows the similarity of the morphology of the two maps.

\begin{figure}[!htb]
  \begin{center}
   \centerline{ \includegraphics[width=0.33\textwidth]{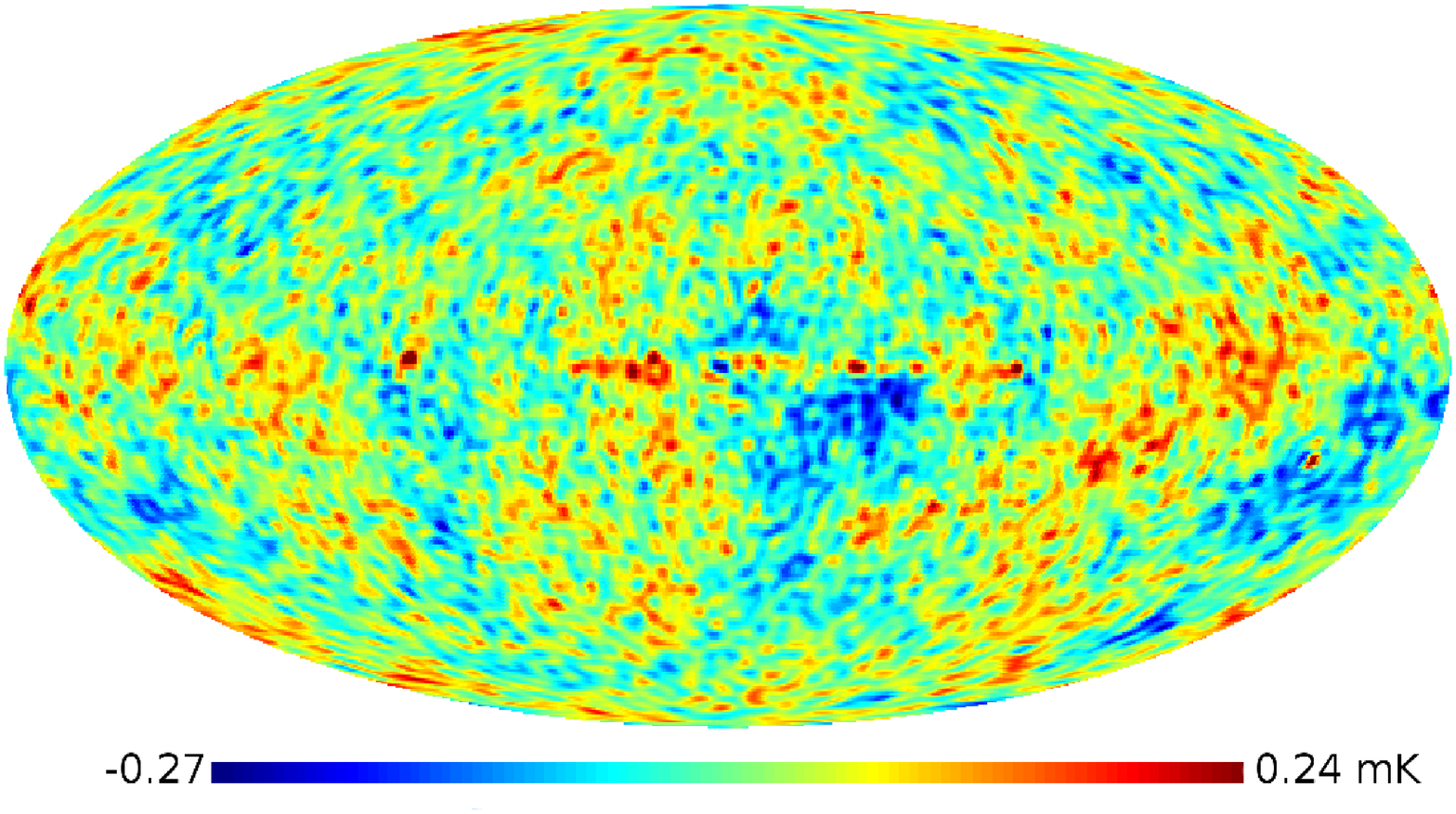}
    \includegraphics[width=0.33\textwidth]{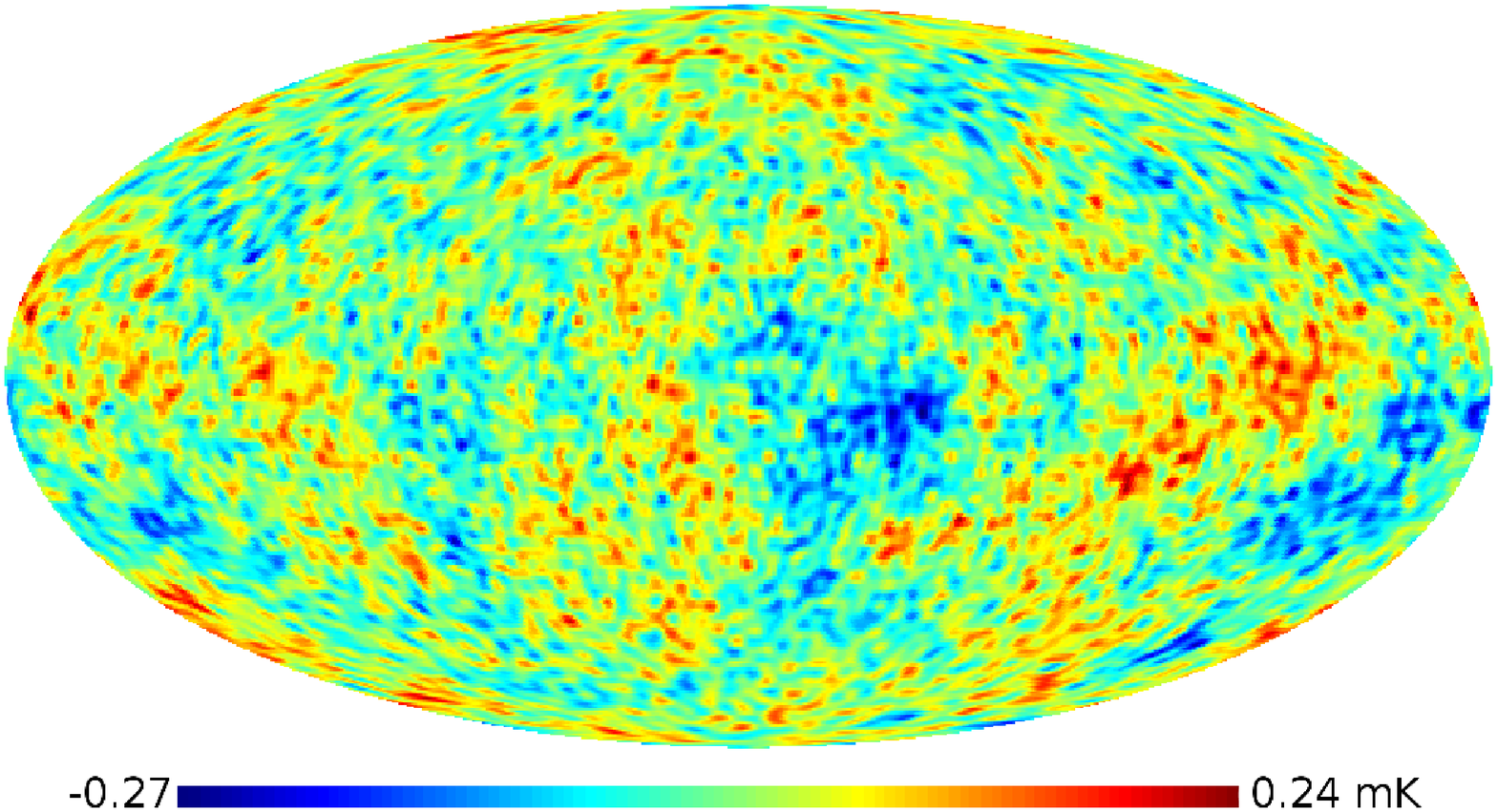}
    \includegraphics[width=0.33\textwidth]{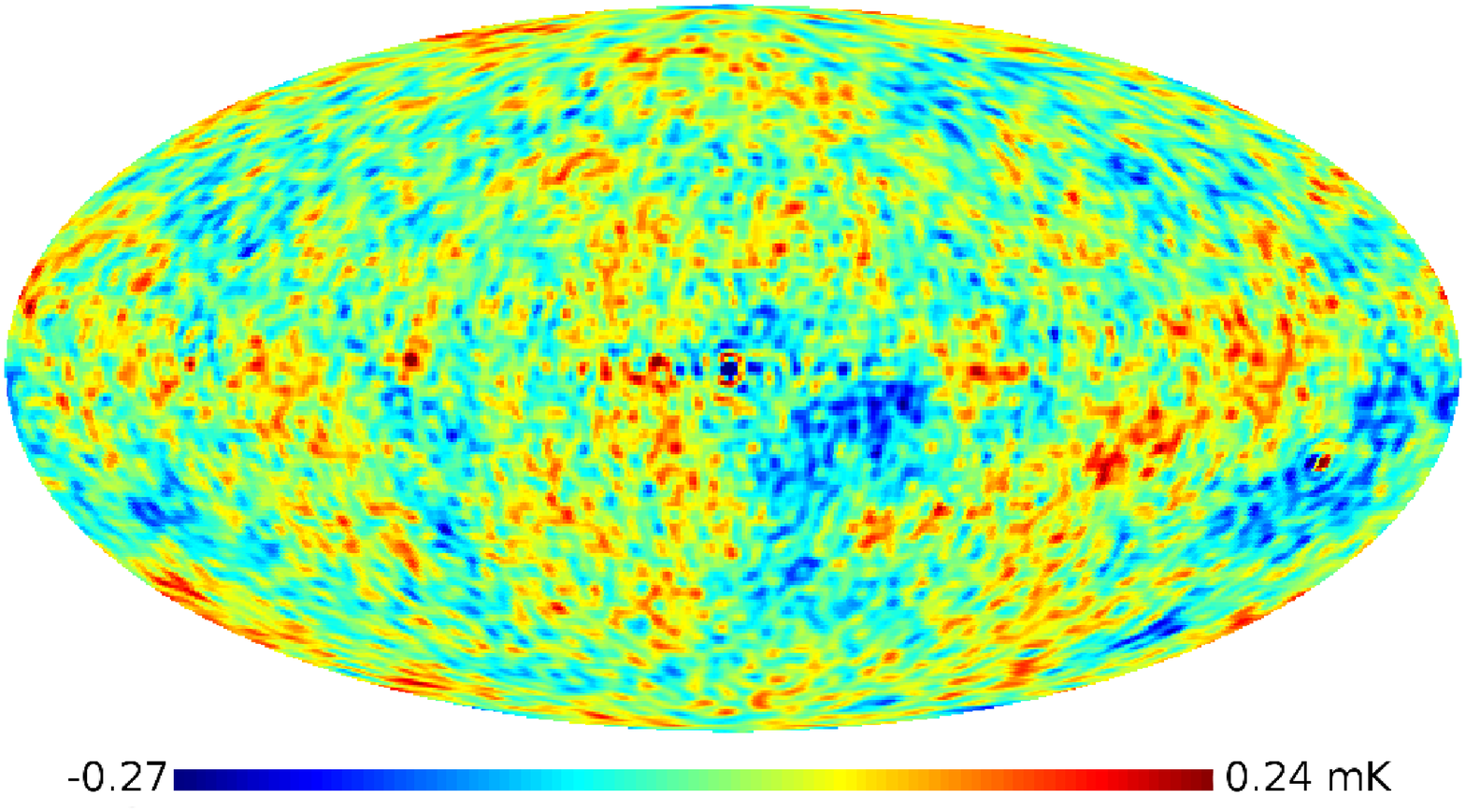}}
    \caption{Left: Planck NILC map. Middle: Planck SMICA map. Right: Planck SEVEM map.}
    \label{fig:planckmaps}
  \end{center}
\end{figure}

\section{Consistency tests}

\subsection{Consistency tests of Planck NILC, SMICA and SEVEM maps}
As mentioned in the introduction, we now perform consistency tests of the three Planck CMB maps: SMICA, NILC and SEVEM (shown in figure \ref{fig:planckmaps}).

\begin{figure}[!ht]
  \begin{center}
       \centerline{\includegraphics[width=0.47\textwidth]{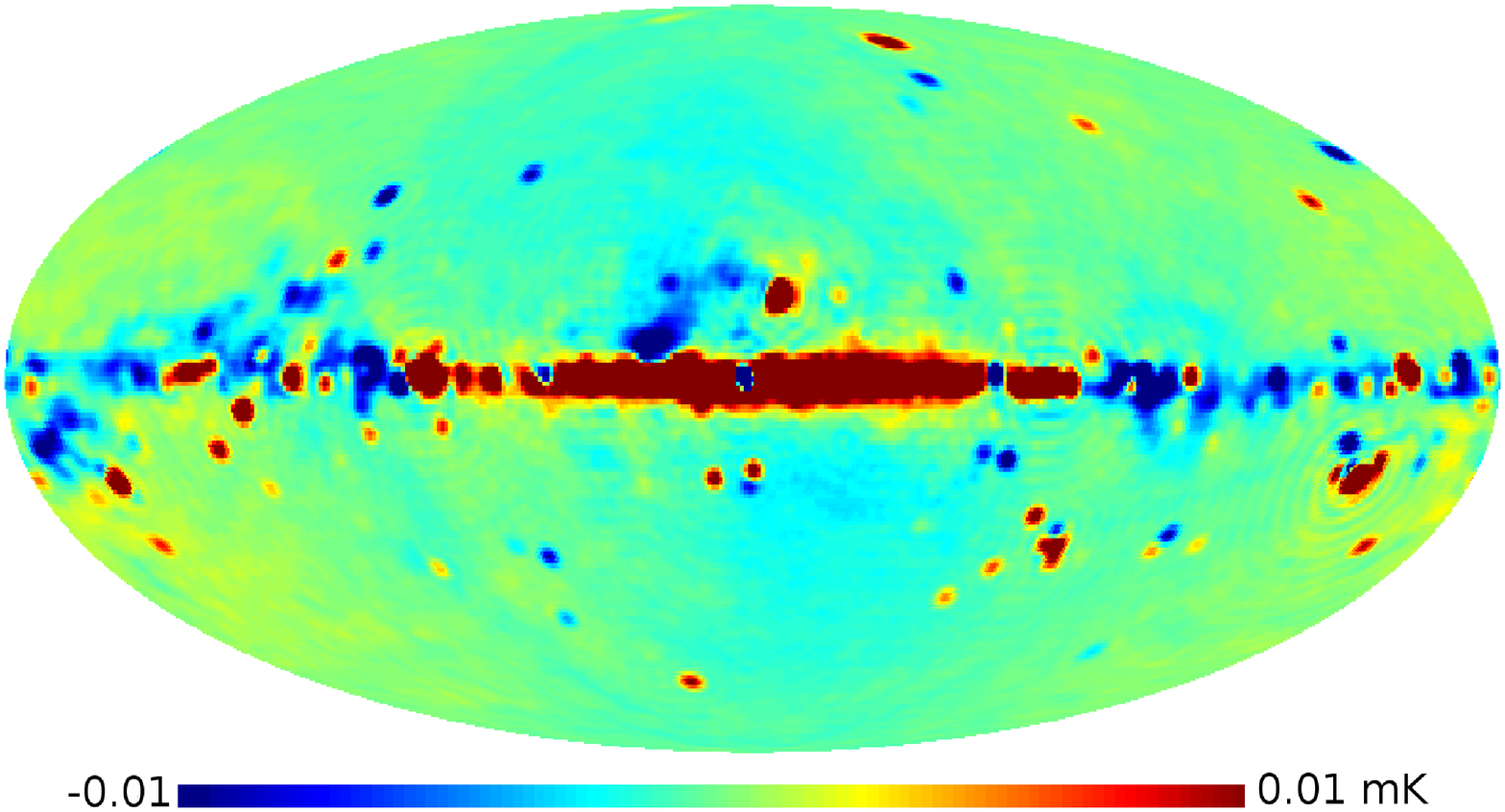}
		   \includegraphics[width=0.47\textwidth]{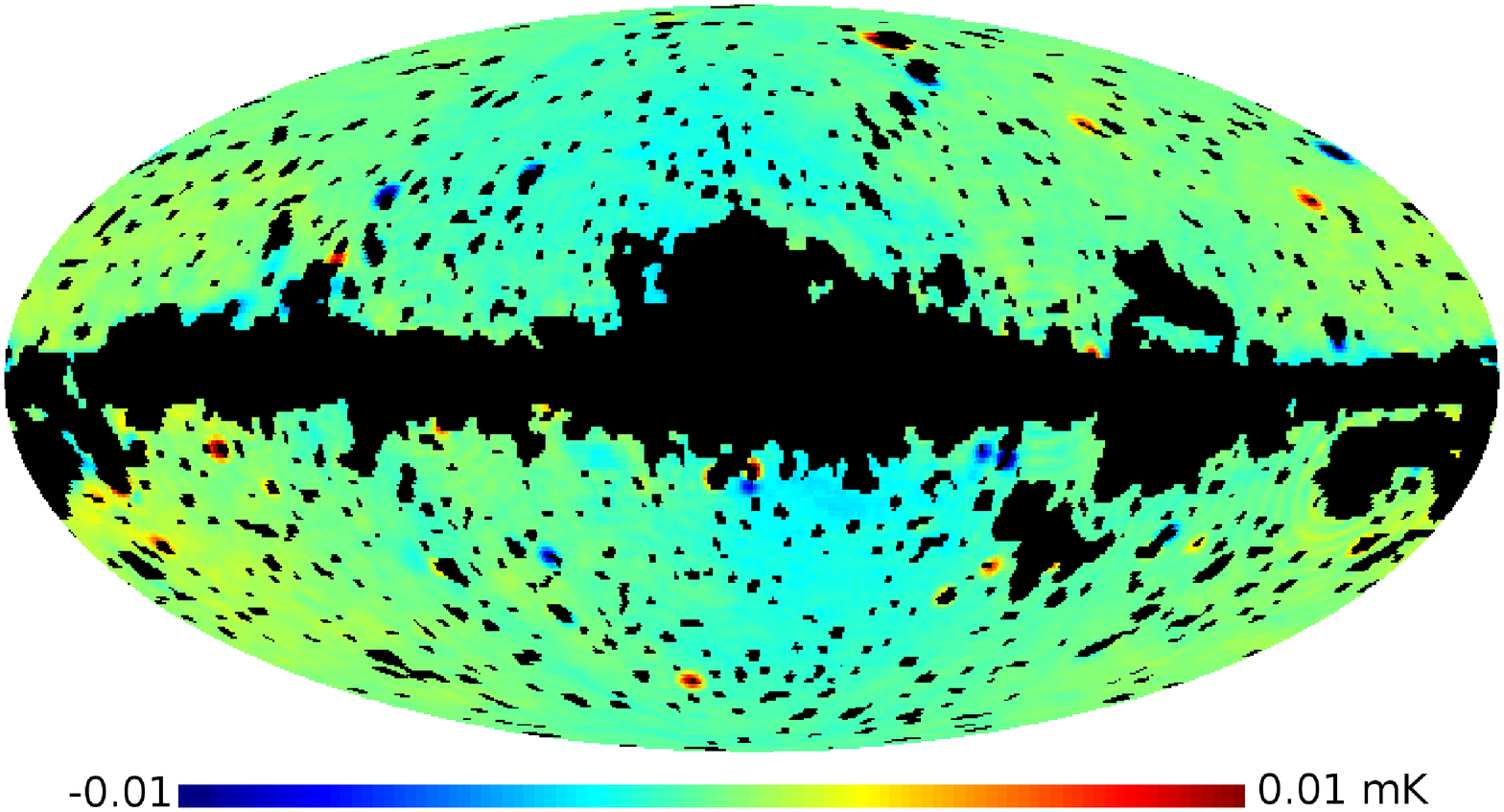}}
	\centerline{\includegraphics[width=0.47\textwidth]{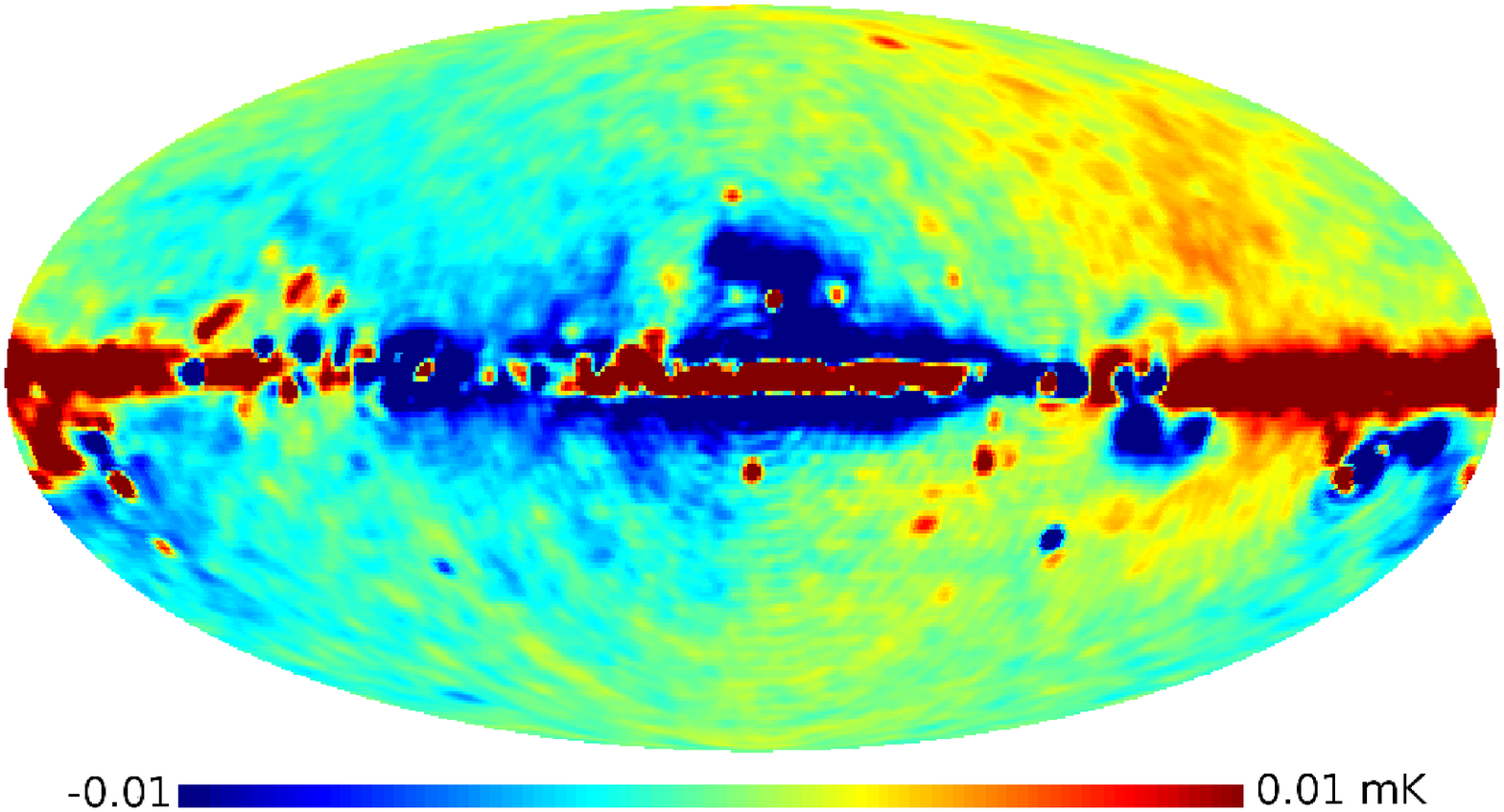}
		   \includegraphics[width=0.47\textwidth]{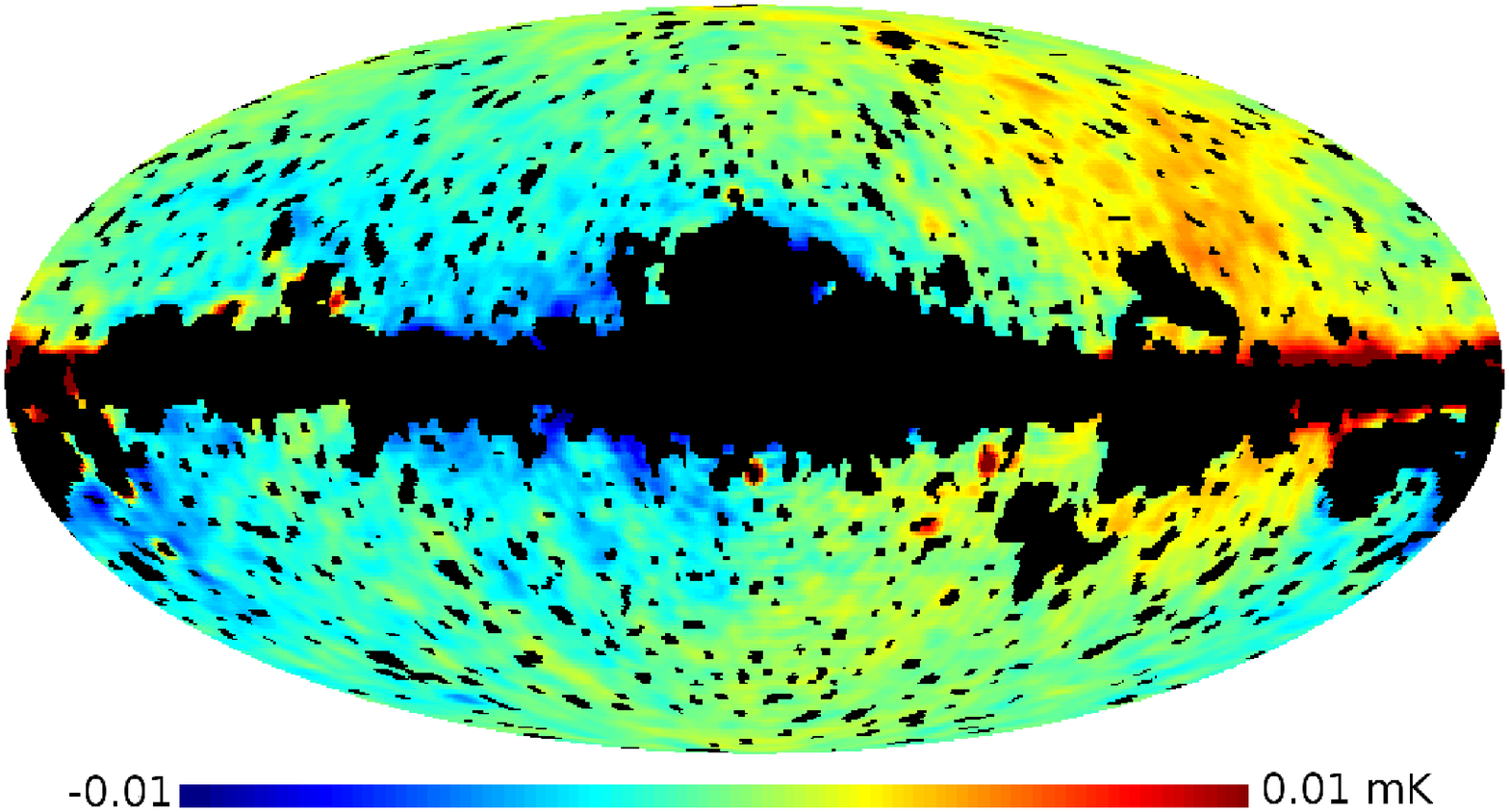}}
	\centerline{\includegraphics[width=0.47\textwidth]{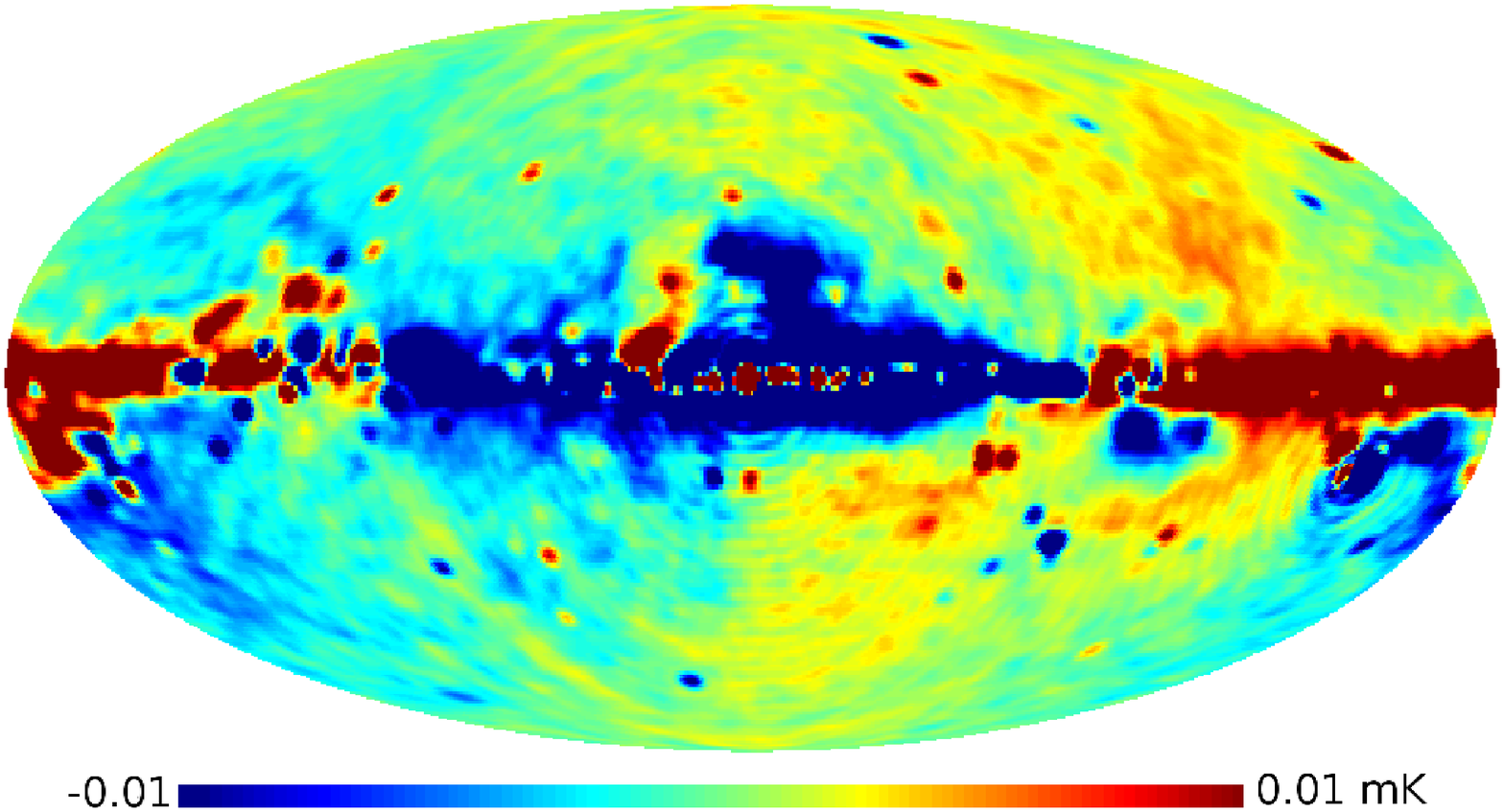}
		   \includegraphics[width=0.47\textwidth]{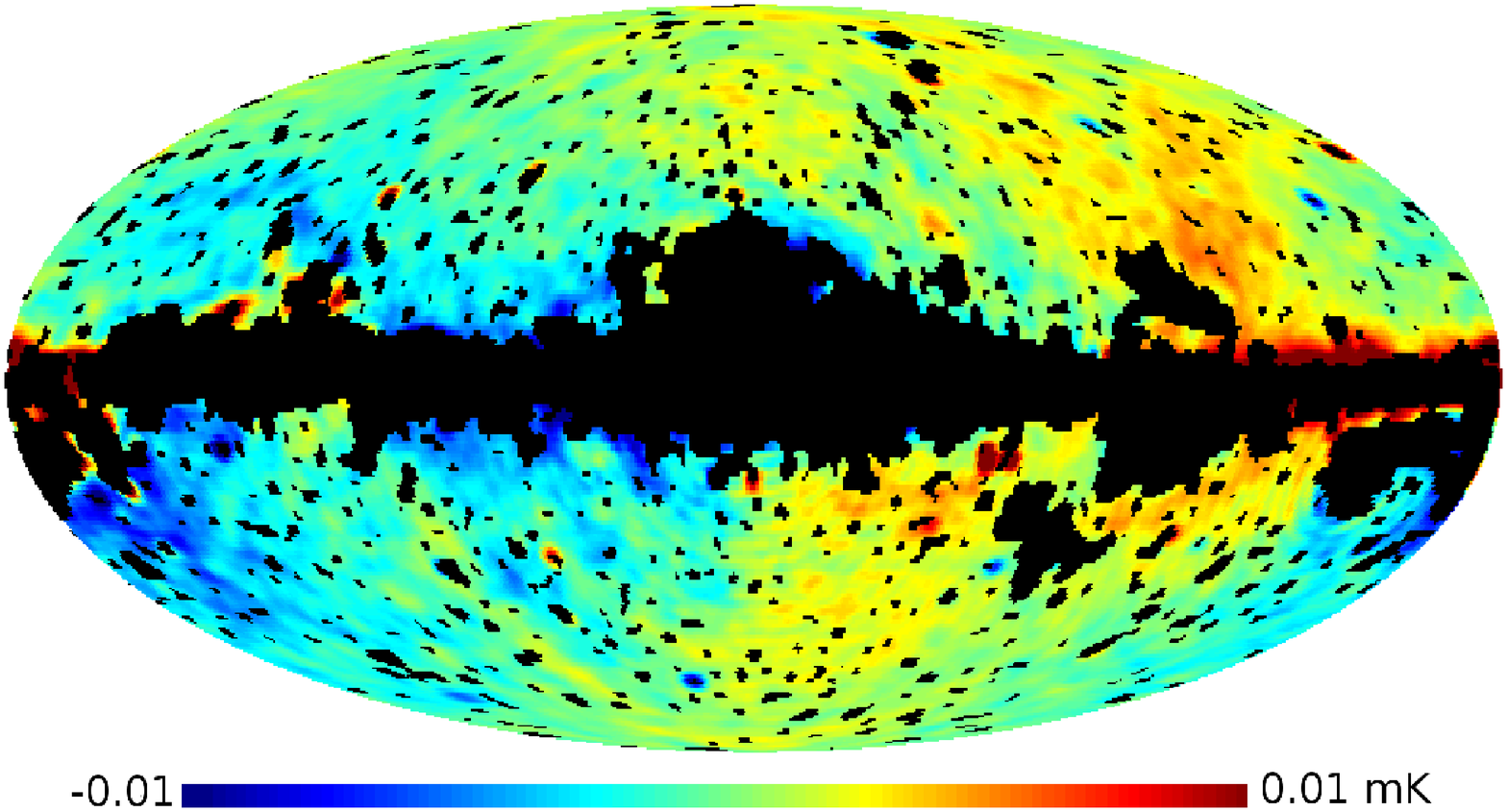}}
    \caption{The difference maps, with the masked map to the right (WMAP KQ85 9yr mask). Top: NILC-SMICA difference map. Middle: NILC-SEVEM difference map. Bottom: SMICA-SEVEM difference map.}
    \label{fig:planckdiffmaps}
  \end{center}
\end{figure}

First, we create the difference map by subtracting the various Planck maps from each other in pixel space. We take three representative combinations of difference maps: NILC minus SMICA (denoted NILC-SMICA), NILC minus SEVEM (NILC-SEVEM) and SMICA minus SEVEM (SMICA-SEVEM). The three difference maps are created with $l_{max} = 100$, and a resolution corresponding to $N_{side} = 128$. In figure \ref{fig:planckdiffmaps} both the unmasked difference maps and the same maps with the KQ85 9yr mask applied are shown for illustrative purposes. Similar maps can be found in \cite{Planck12}, albeit with a different color scheme and other temperature bounds. 

\begin{figure}[!ht]
  \begin{center}
        \centerline{\includegraphics[width=0.35\textwidth]{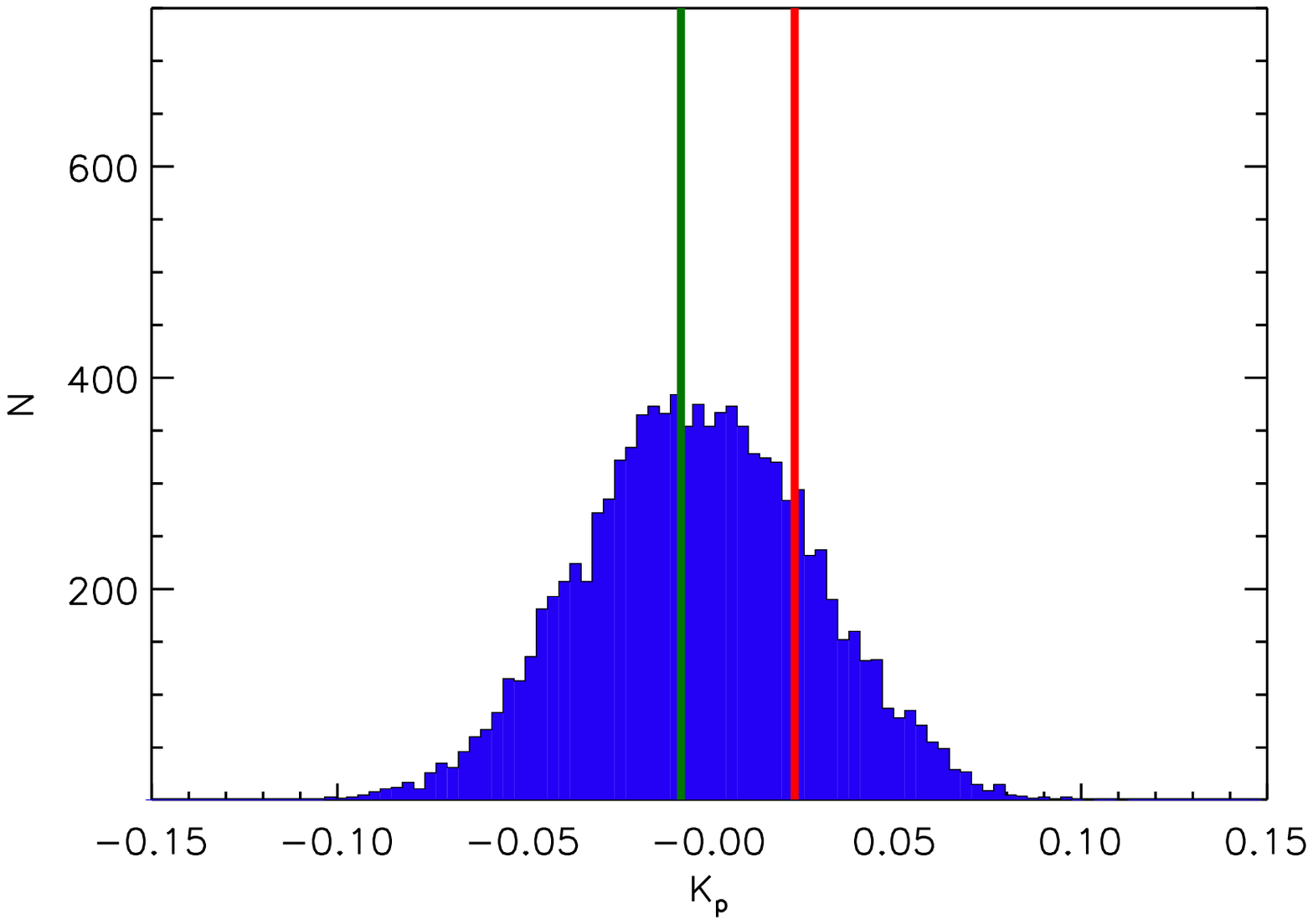}
    \includegraphics[width=0.35\textwidth]{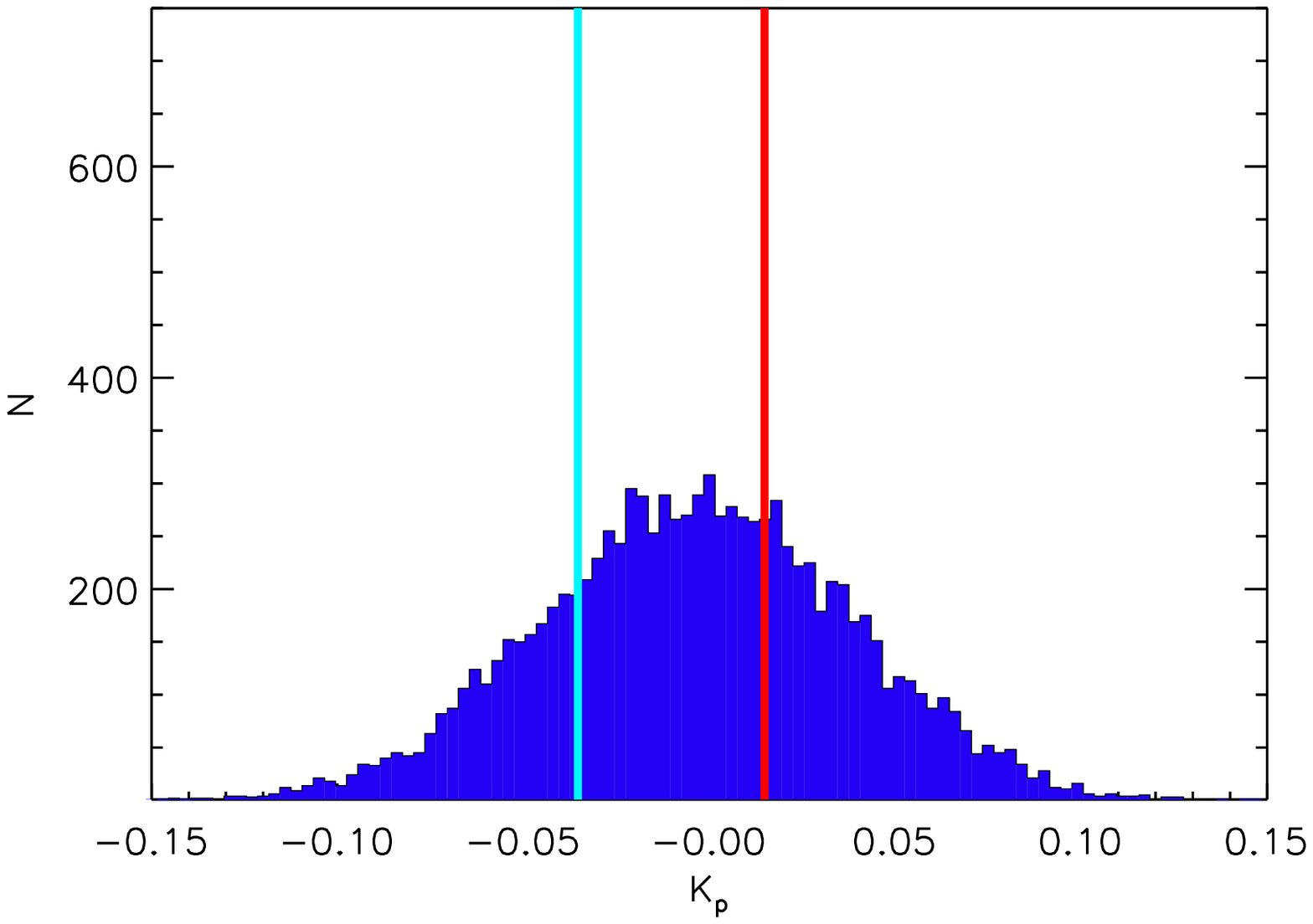}      
    \includegraphics[width=0.35\textwidth]{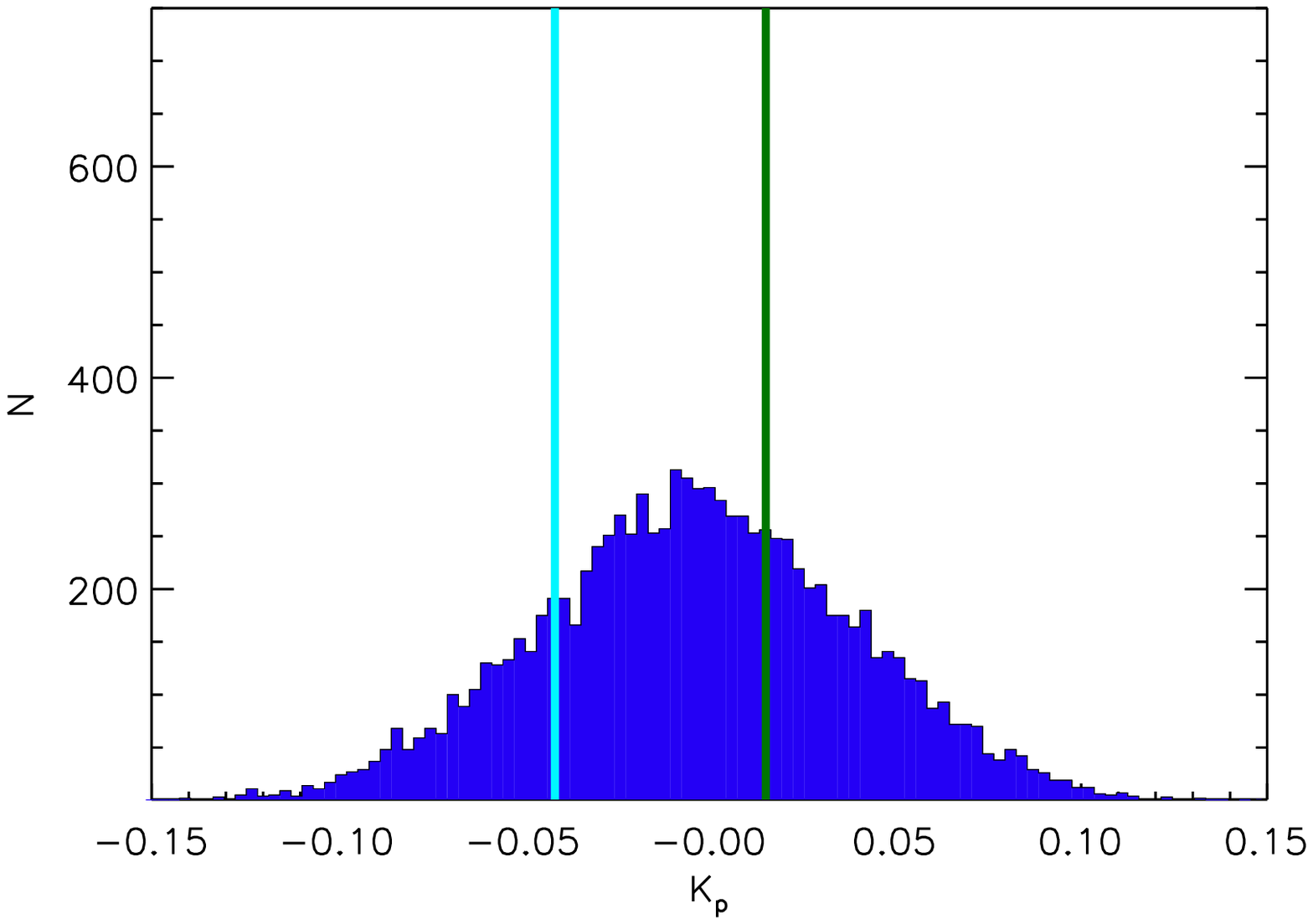}}
    \caption{Cross correlations coefficients compared to 10000 simulated maps. Top left is for the NILC-SMICA, top right is for NILC-SEVEM and bottom is for SMICA-SEVEM. Red line: NILC, green line: SMICA, light blue line: SEVEM. All with the respective difference map.}
    \label{fig:hist_pix_crosscorr}
  \end{center}
\end{figure}

Following to Eq. (\ref{eq:crossp}) we now calculate $K_p$, where we cross correlate each input map with the difference map. We compare this to the result for 10000 random simulations, in order to asses the significance. The results are presented in figure \ref{fig:hist_pix_crosscorr}. We see that the Planck maps only correlate weakly with their respective difference maps, and are well consistent with simulations. We attribute this to the high similarity between the Planck maps (see figure \ref{fig:planckdiffmaps}) outside the Galactic mask, making the numerator in Eq. (\ref{eq:crossp}) small. In conclusion, the three Planck maps are very consistent with each other. The numerical values are shown in table \ref{tab:planck1}. 

\begin{table}[!ht]
\centering
  \begin{tabular}{l | r | c | r | c | r | c|}
\cline{2-7}\cline{2-7}
			& \multicolumn{2}{c|}{\textbf{NILC-SMICA}} 	& \multicolumn{2}{c|}{\textbf{NILC-SEVEM}} 	& \multicolumn{2}{c|}{\textbf{SMICA-SEVEM}} 	\\ \cline{2-7}
			& $K_p$  \phantom{0}\phantom{0}	& Percentage	& $K_p$ \phantom{0}\phantom{0} 	& Percentage	& $K_p$  \phantom{0}\phantom{0}		& Percentage	\\ \hline \hline 
 \multicolumn{1}{|c|}{\textbf{NILC}}& 0.02294\phantom{0}& 23.4\%	& 0.01485 		& 36.0\%	&\cellcolor[gray]{0.9}			&\cellcolor[gray]{0.9}		\\ \hline 
 \multicolumn{1}{|c|}{\textbf{SMICA}}& -0.007667 	& 41.0\%	& \cellcolor[gray]{0.9}&\cellcolor[gray]{0.9}	& 0.01518				& 35.4\%		\\ \hline 
 \multicolumn{1}{|c|}{\textbf{SEVEM}}& \cellcolor[gray]{0.9}&\cellcolor[gray]{0.9}& -0.03536		& 20.6\%	& -0.04156				& 16.7\%		\\ 
  \hline \hline
  \end{tabular} 
\label{tab:planck1}
\caption{Numerical values of the cross correlations for figure \ref{fig:hist_pix_crosscorr}. For each difference map, the table shows the value of $K_p$ for a map cross correlated with the respective difference map, and the percentage of the 10000 simulations that have a higher (or for $K_p<0$: lower) value of $K_p$.}
\end{table}

\begin{figure}[!ht]
  \begin{center}
     \centerline{\includegraphics[width=0.47\textwidth]{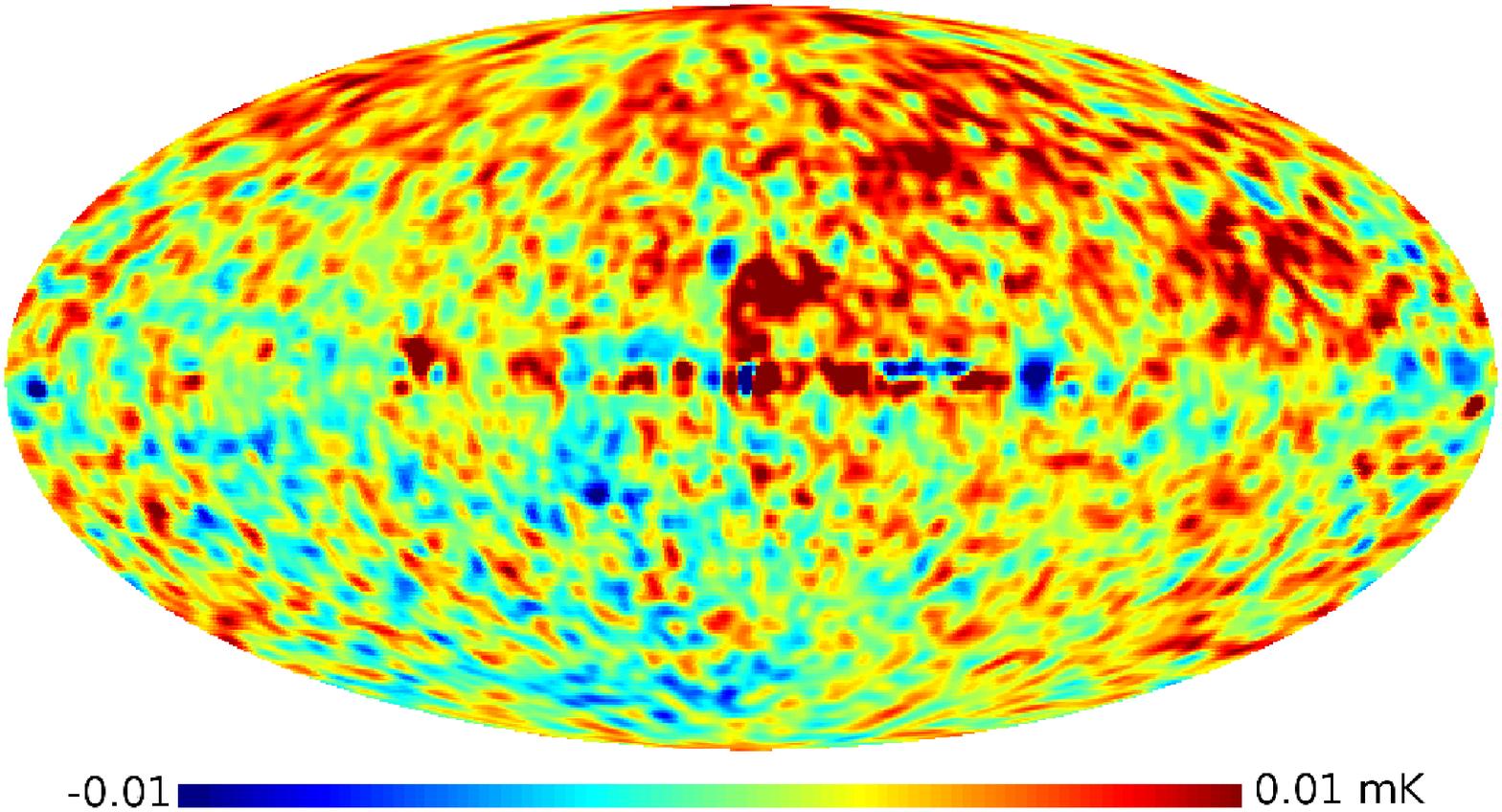}
		  \includegraphics[width=0.47\textwidth]{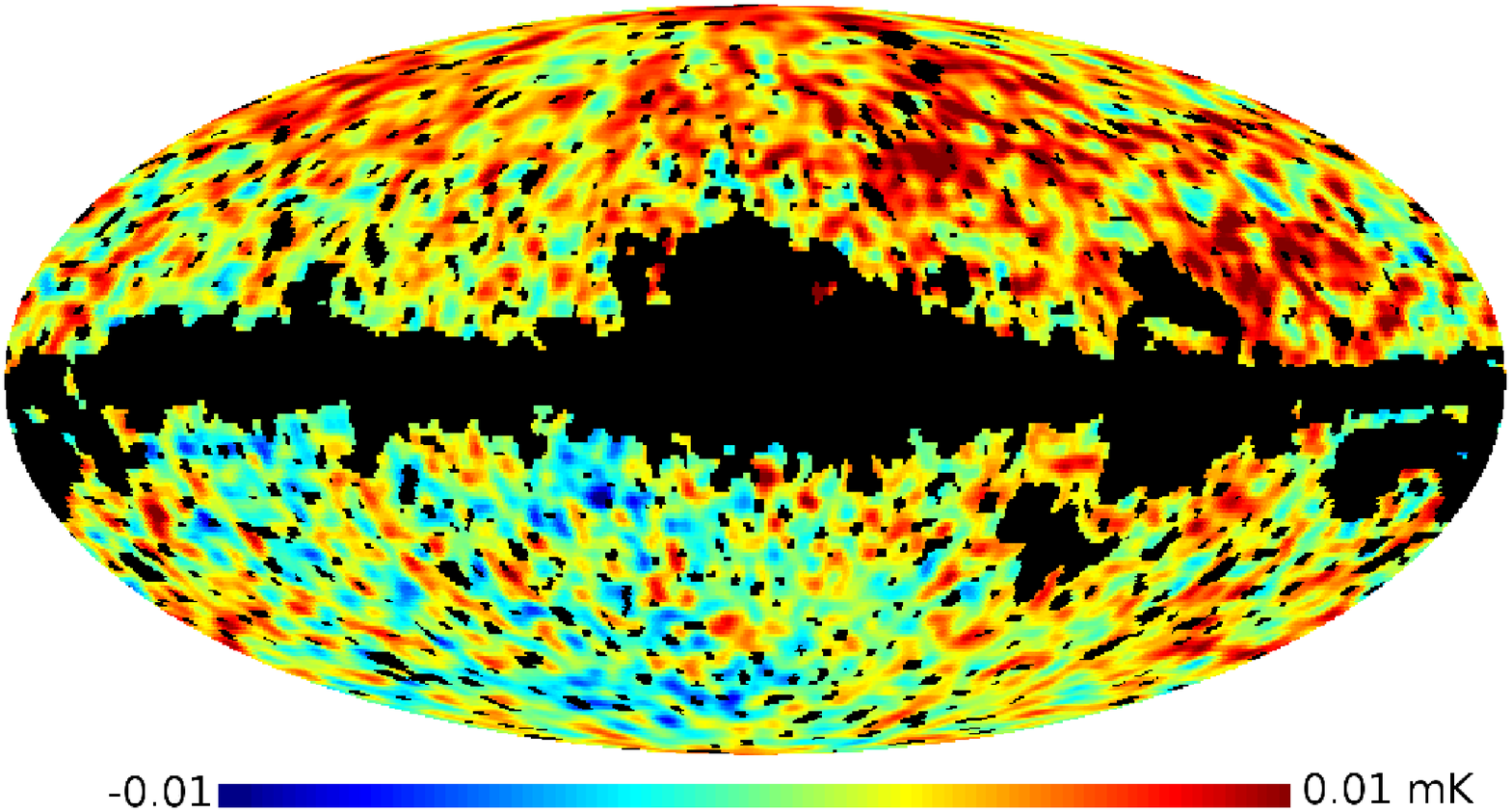}}
		 \includegraphics[width=0.47\textwidth]{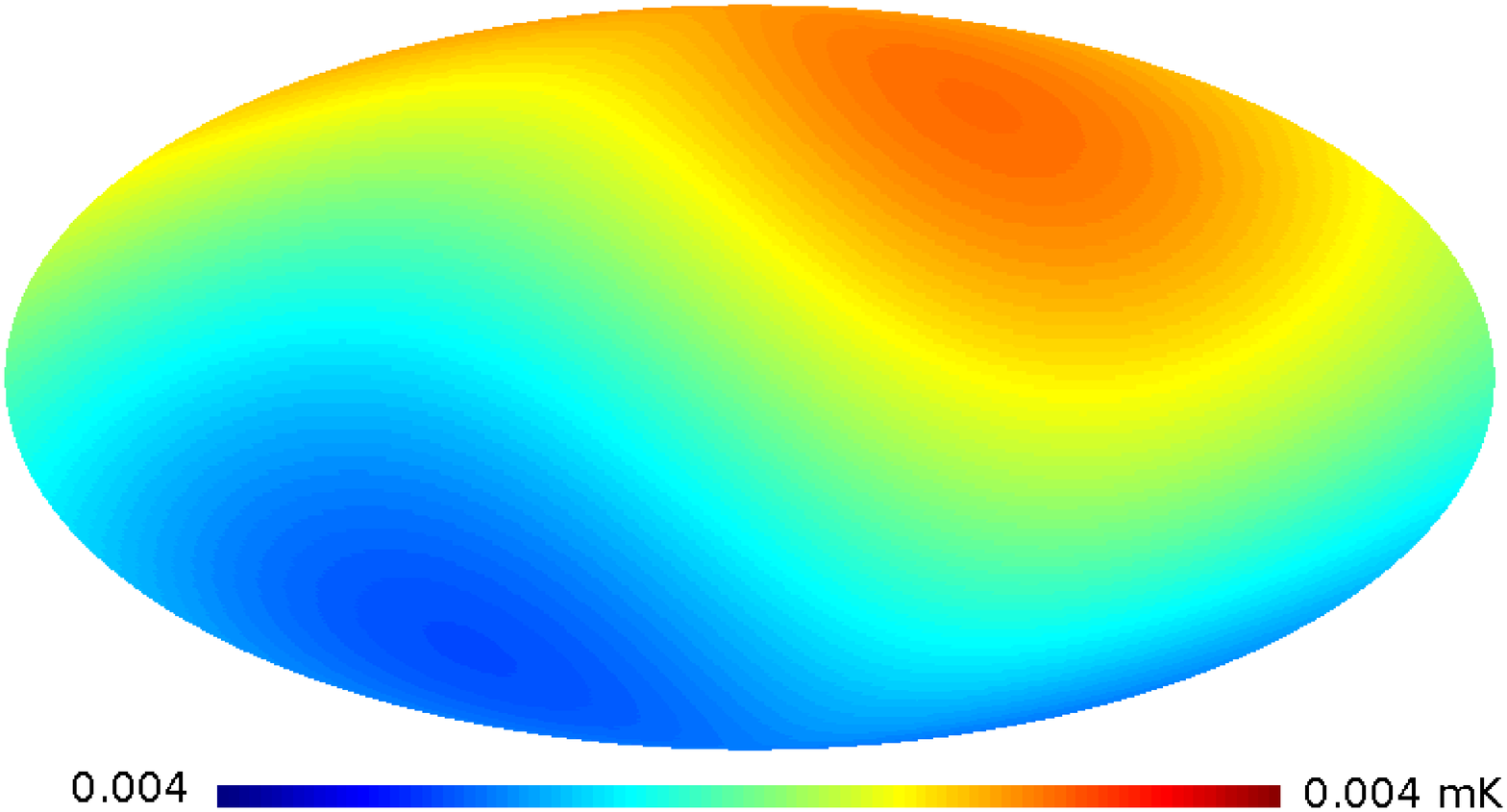}
     \caption{Top left: The difference map, computed as ILC9-ILC7. Top right: the masked difference map (WMAP KQ85 9yr mask). Bottom: Dipole ($l=1$) of the difference map ILC9-7.}
    \label{fig:diffmap}
  \end{center}
\end{figure}

\subsection{Consistency tests of WMAP ILC9 and ILC7 maps}
The weights in the construction of the ILC9 have been improved from the ILC7 through refinement of the pixel noise, calibrations and beam profiles (see \cite{WMAP9} for details). The ILC9 map is thus expected to be superior to the ILC7 through 2 additional years of data taking as well as optimization of the method of construction. 

\begin{figure}[!ht]
\begin{floatrow}
\ffigbox[0.5\textwidth]
{\includegraphics[width=0.45\textwidth]{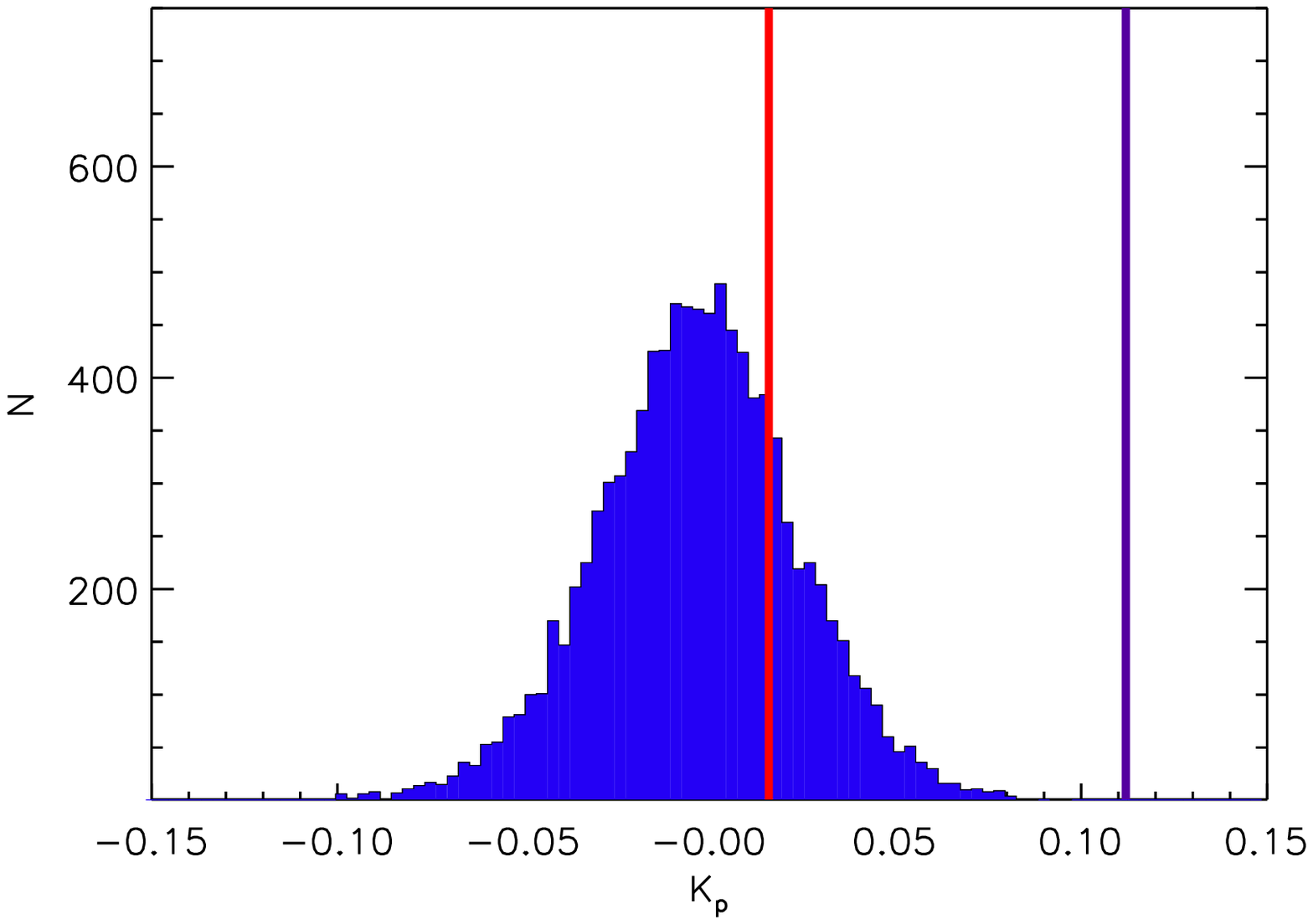}}
{\caption{Cross correlations coefficients compared to 10000 simulated maps for the ILC9-7 difference map. Red: ILC7. Purple: ILC9.}
\label{fig:ilc9-7_corr}}
\capbtabbox[0.45\textwidth]{\qquad
 \begin{tabular}{l | r | c |}
\cline{2-3}\cline{2-3}
					& \multicolumn{2}{c|}{\textbf{ILC9-7}} 	\\ \cline{2-3}
					& $K_p$  \phantom{0}\phantom{0}	& Percentage	\\ \hline \hline
   \multicolumn{1}{|c|}{\textbf{ILC9}}	& 0.1120\phantom{0} 		& $<0.01\%$	\\ \hline 
   \multicolumn{1}{|c|}{\textbf{ILC7}} 	& 0.01597			& 26.5\%	\\ 
  \hline \hline
  \end{tabular} 
\vspace{1.5cm}
}{\qquad
  \caption{Numerical values for figure \ref{fig:ilc9-7_corr}. The table shows the value of $K_p$ for a map cross correlated with the respective difference map (left), and the percentage of the 10000 simulations with a higher value of $K_p$.}
\label{tab:wmap}
}
\end{floatrow}
\end{figure}

Similar to the test for the individual Planck maps, we therefore now turn our attention to the ILC9 in comparison with the ILC7. The ILC9-7 difference map is shown in figure \ref{fig:diffmap}. The galactic plane is clearly visible in the ILC9-7 difference map, as are selected point sources. Therefore we mask it with the KQ85 9yr mask, as we did in the case for the Planck maps (figure \ref{fig:diffmap}, top right). It is immediately clear that the map is dominated by a dipole (see bottom figure), which is closely aligned with the well known kinematic dipole. The same feature is present in the difference map between the 7 year and 5 year ILC map, as discovered in \cite{WMAP7-1}.  

We compute the cross correlation between the ILC9 and ILC7 year maps with the ILC9-7 difference map. The procedure is the same as in the previous section. The results are presented in figure \ref{fig:ilc9-7_corr}, and the numerical values for the cross correlation are presented in table \ref{tab:wmap}. We see that ILC9 cross correlates much stronger with the difference map than ILC7, and  that the ILC7 is in agreement with the distribution of the 10000 random simulations, while the ILC9 is not.

\subsection{Consistency tests of Planck NILC and WMAP ILC9 maps}
\begin{figure}[!ht]
  \begin{center}
     \centerline{\includegraphics[width=0.5\textwidth]{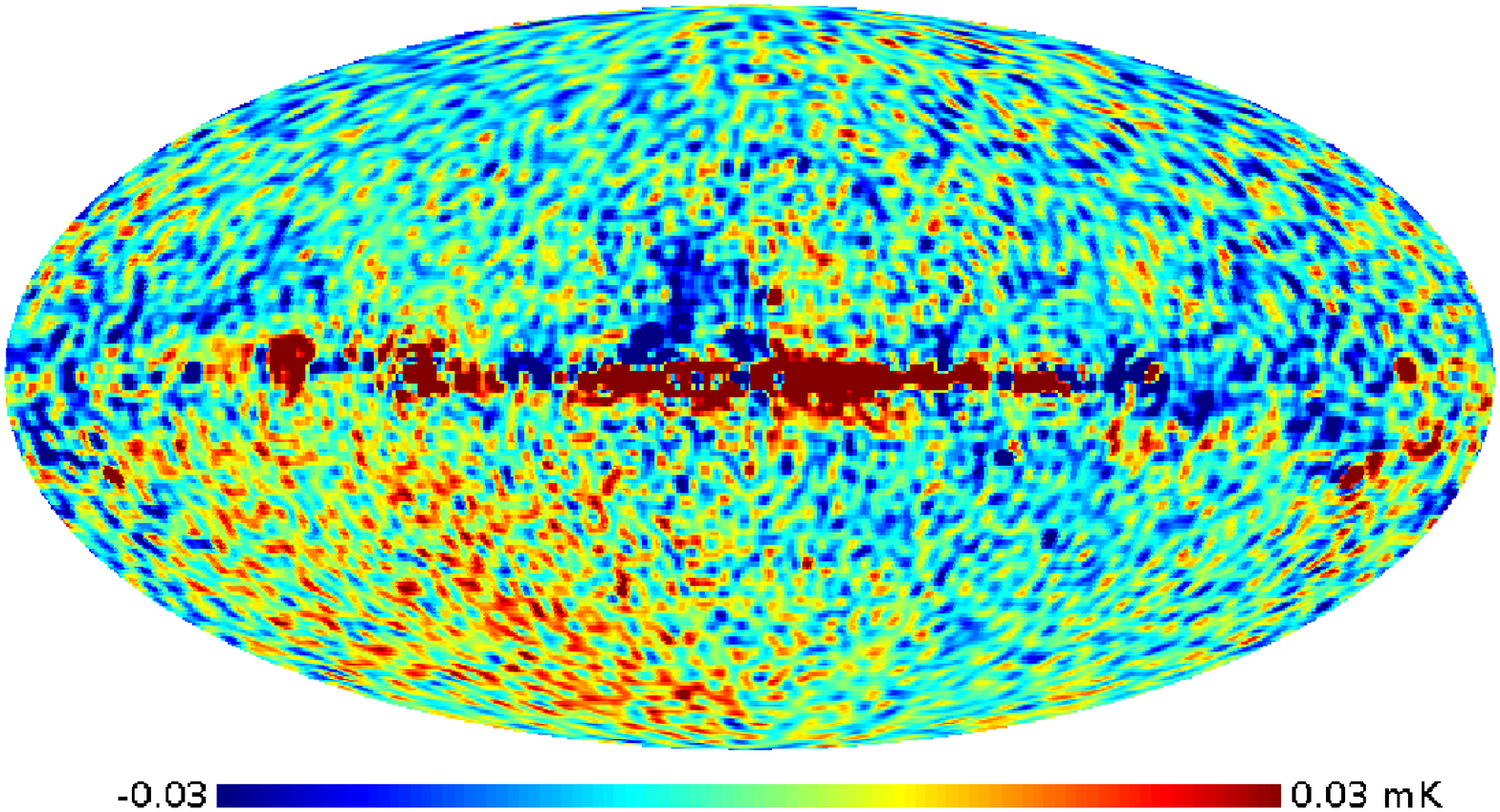}
		  \includegraphics[width=0.5\textwidth]{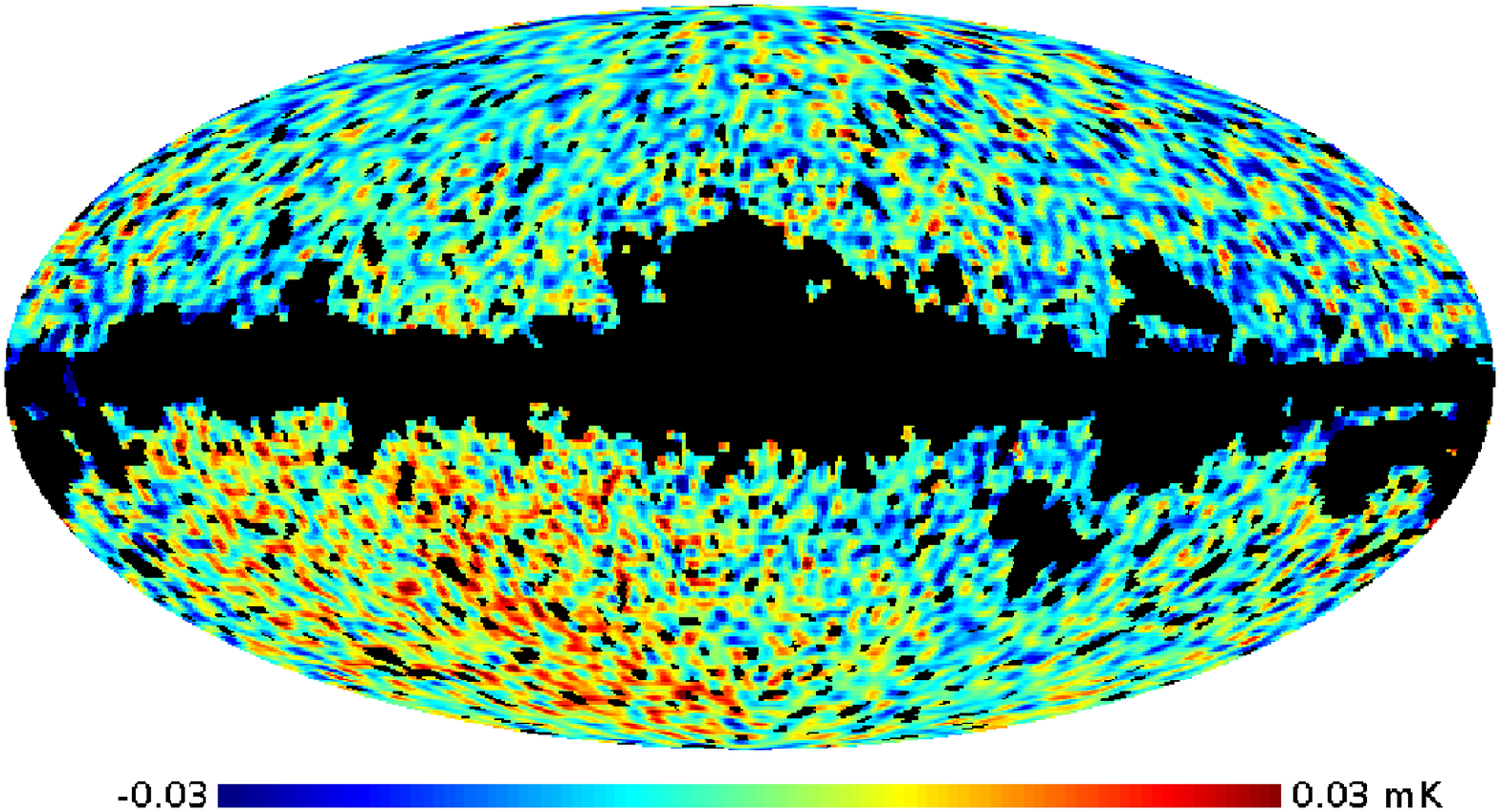}}
     \caption{Left: the difference map, computed as NILC-ILC. Right: the masked difference map (WMAP KQ85 9yr mask). }
    \label{fig:diffmap_planck_wmap}
  \end{center}
\end{figure}

Finally, we turn to an investigation of the difference map between the Planck NILC map and the WMAP ILC9 map. We select the Planck NILC map for the comparison, since the NILC method is similar in nature to the ILC method (see for instance \cite{Planck12}). Since the WMAP ILC9 map has been smoothed at the level of $1^\circ$, we start by smoothing the NILC at the same level. Then the difference map is calculated similar to the procedure in the previous sections, and the result is shown in figure \ref{fig:diffmap_planck_wmap}. We have enhanced the min and max temperature compared to previous difference maps in the paper, in order to clearly show local features. In the difference map we clearly see the galactic plane, and some feature in the lower left quadrant of the map. This is similar to the dipole clearly seen in figure \ref{fig:diffmap}, but since we are now subtracting the ILC9 map, the sign of the dipole is changed. Is is evident that the ILC9 contains an enhanced dipole, both in comparison with the WMAP ILC7 and with the Planck NILC map.

\begin{figure}[ht]
\begin{floatrow}
\ffigbox[0.5\textwidth]
{\includegraphics[width=0.5\textwidth]{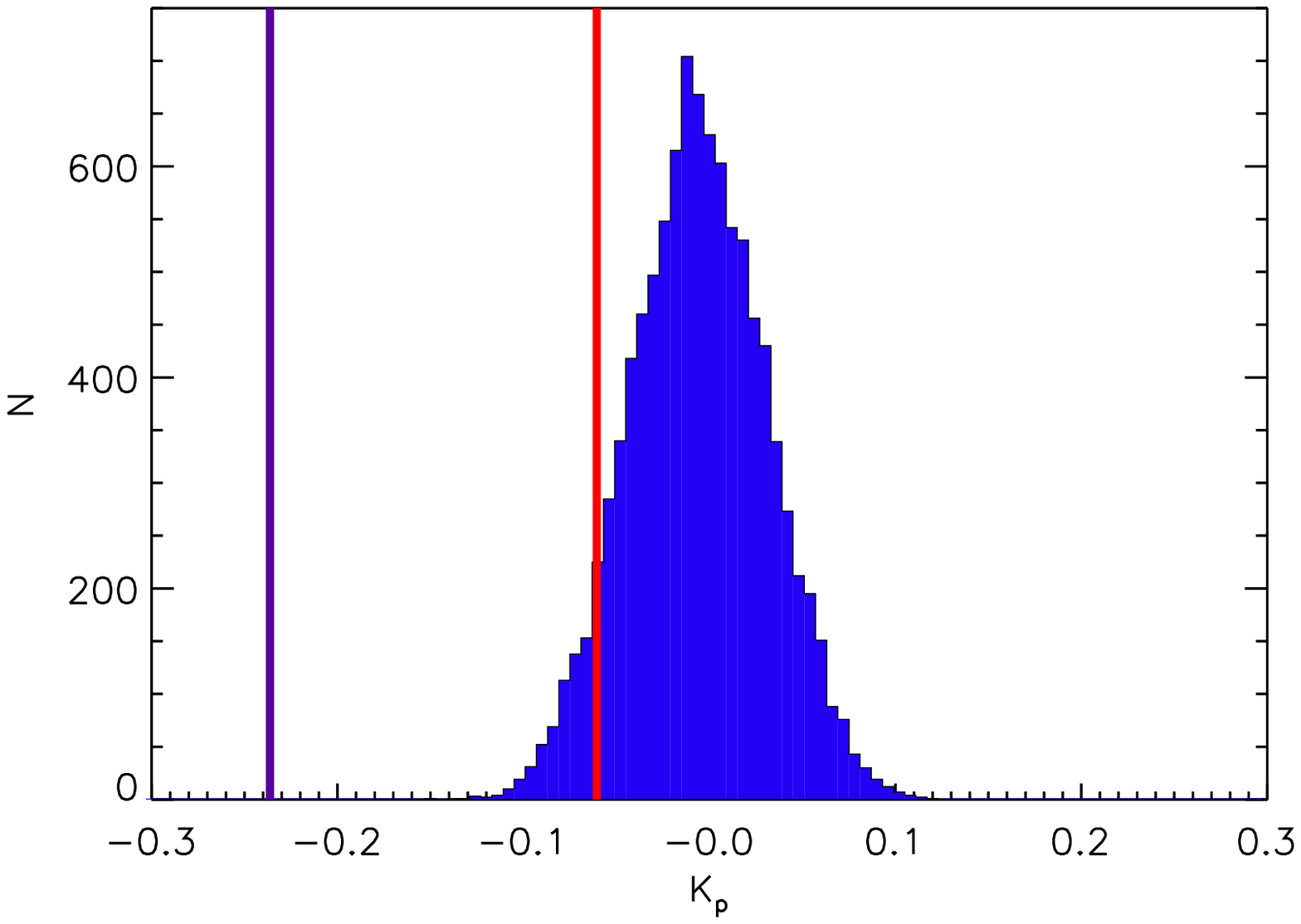}}
{\caption{Cross correlation coefficients compared to 10000 simulated maps for the NILC-ILC9 difference map. Red: NILC. Purple: ILC9.}
\label{fig:hist_pix_nilc_ilc9}}
\capbtabbox[0.45\textwidth]{\qquad
 \begin{tabular}{l | c | c |}
\cline{2-3}\cline{2-3}
					& \multicolumn{2}{c|}{\textbf{NILC-ILC9}} 	\\ \cline{2-3}
					& $K_p$				& Percentage	\\ \hline \hline
   \multicolumn{1}{|c|}{\textbf{NILC}}	& -0.0606 		& \phantom{0}94.76\%	\\ \hline 
   \multicolumn{1}{|c|}{\textbf{ILC9}} 	& -0.2362			& $<0.01\%$	\\ 
  \hline \hline
  \end{tabular} 
\vspace{1.5cm}
}{\qquad
  \caption{Numerical values for figure \ref{fig:hist_pix_nilc_ilc9}. The table shows the value of $K_p$ for a map cross correlated with the respective difference map (left), and the percentage of the 10000 simulations with a higher value of $K_p$.}
\label{tab:Planck-WMAP}
}
\end{floatrow}
\end{figure}

In figure \ref{fig:hist_pix_nilc_ilc9} we show the result of the cross correlations for the difference map and the NILC (smoothed to $1^\circ$) and ILC9 map respectively, in comparison to 10000 simulated maps. Numerical values are presented in table \ref{tab:Planck-WMAP}. The resolution, mask and maximum $l$-value is similar to the previous tests.
We see a strong (anti-)correlation for the ILC9 cross correlated with NILC-ILC9, in comparison with the simulations. The cross correlation case for the NILC is in reasonable accordance with the simulations. In comparison with the case for ILC9-7 in the previous section this indicates that the ILC9 map includes a proportion of non-cosmological features.

\section{Summary and conclusions}
In this paper we have investigated the consistency of the Planck NILC, SMICA and SEVEM maps as well as the consistency of the WMAP ILC9 and ILC7 maps, and a cross test between NILC and ILC9. The basic assumption was that the difference between two maps--the difference map--consists only of a non-cosmological signal, which a pure CMB map should not correlate strongly with. We have tested this through comparing pairs of maps to their respective difference map, and the significance of the correlation was determined through comparison with 10000 random simulations of the CMB. 

For the consistency test of the derived Planck products, we cross correlated the three Planck maps with the respective difference maps NILC-SMICA, NILC-SEVEM and SMICA-SEVEM at $N_{side}=128$ (all masked with the KQ85 9 year mask). We found that the Planck maps are very well consistent with each other, as well as consistent with random simulations. 

Additionally, we find that the ILC9-ILC7 difference map cross correlates significantly with the ILC9 map, whereas the ILC7 cross correlation is consistent with the random simulations (both at $N_{side}=128$). This shows that the ILC9 map matches the morphology of the non-cosmological contributions in the difference map, and the ILC7 map does not significantly.

Finally, we did an external consistency test, comparing the Planck NILC map with the ILC9 map. The difference map between NILC and ILC9 showed a dipole, also visible for the ILC9-7 case, indicating that this is probably an artifact of the ILC9 map. Also, the NILC-ILC9 difference map cross correlates strongly with the ILC9 and much less so with the NILC map.

In conclusion, the Planck NILC, SEVEM and SMICA maps are in very good agreement with each other, and none of them show a significant correlation with their respective difference maps outside the mask. On the other hand, the ILC9 map correlates significantly with the difference map, both in comparison to ILC7, in comparison to Planck NILC and in comparison to random simulations. Thus ILC9 appears to be more contaminated than the ILC7. This should be taken into consideration when using WMAP maps for cosmological analyses. 

\section*{Acknowledgements}
We are grateful for discussions on this work with P. Naselsky, K. G\'{o}rski, C. Burigana, J. Tauber, C. Lawrence, M. Bersanelli, P. Natoli and A. Bandy. 

This work is based on observations obtained with Planck ({\fontsize{9.6}{6}\selectfont \url{http://www.esa.int/Planck}}), an ESA science mission with instruments and contributions directly funded by ESA Member States, NASA, and Canada. The development of Planck has been supported by: ESA; CNES and CNRS / INSU-IN2P3-INP (France); ASI, CNR, and INAF (Italy); NASA and DoE (USA); STFC and UKSA (UK); CSIC, MICINN and JA (Spain); Tekes, AoF and CSC (Finland); DLR and MPG (Germany); CSA (Canada); DTU Space  (Denmark); SER/SSO (Switzerland); RCN (Norway); SFI (Ireland); FCT/MCTES (Portugal); and PRACE (EU).
A description of the Planck Collaboration and a list of its members, including the technical or scientific activities in which they have been involved, can be found at the Planck web page\footnote{\url{http://www.sciops.esa.int/index.php?project=planck&page=Planck_Collaboration}}.

We acknowledge the use of the NASA Legacy Archive for Microwave Background Data Analysis (LAMBDA). Our data analysis made use of the GLESP package \cite{Glesp}\footnote{\url{http://www.glesp.nbi.dk/}}, and of HEALPix \cite{Healpix_primer}. This work is supported in part by Danmarks Grundforskningsfond which allowed the establishment of the Danish Discovery Center, FNU grants 272-06-0417, 272-07-0528 and 21-04-0355, the National Natural Science Foundation of China (Grant No. 11033003), the National Natural Science Foundation for Young Scientists of China (Grant No. 11203024) and the Youth Innovation Promotion Association, CAS.

\end{document}